\documentclass[12pt]{article}

\newcommand{\p}{\partial}

\begin{document}
\topmargin 0pt
\oddsidemargin 0mm

\renewcommand{\thefootnote}{\fnsymbol{footnote}}
\begin{titlepage}
\begin{flushright}IC/2001-55\\
hep-th/0106198
\end{flushright}

\vspace{5mm}
\begin{center}
{\Large \bf On Lie point symmetry of classical Wess-Zumino-Witten Model}
\vspace{10mm}

{\large
Karmadeva Maharana$^{*, \dag}$ \\
\vspace{4mm}

\footnote{Present address}{\em Abdus Salam International Centre for Theoretical Physics, \\
Strada Costiera 11, 34014 Trieste, Italy \\
and \\
\footnote{Permanent address}Physics Department, Utkal University, Bhubaneswar 
751 004, India}\\ 

\vspace{5mm}

email: karmadev@iopb.res.in}

\end{center}
\vspace{5mm}
\centerline{{\bf{Abstract}}}
\vspace{5mm}

We perform the group analysis of Witten's equations of motion for a
particle moving in the presence of a magnetic monopole, 
and also when constrained to move on the surface of a sphere, which is 
the classical example of Wess-Zumino-Witten model. We also 
consider variations of this model. Our analysis
gives the generators of the corresponding Lie point symmetries. 
The Lie symmetry corresponding to Kepler's third law is obtained in 
two related examples. 
 
\end{titlepage}

\newpage

\section{Introduction}

The Wess-Zumino model\cite{wess} has the distiction of being applicable
 via the 
 current algebra conditions to explain the process  $ {{K^{+}} {K^{-}}} {\rightarrow}
{ {{\pi}^{+}}  {{\pi}^{0}}
{{\pi}^{-}}} $
as well as in setting up the string theory actions through the generalizations
made by Witten\cite{witten}. To motivate before arriving at the generalized 
form, Witten
had considered an example of the equation of motion of a particle in three
dimensions constrained to move on the surface of a sphere in the presence
of a magnetic monopole. The equations of motion of such a system in the
presence of a magnetic monopole can not be obtained from the usual Lagrangian
formulation. Hence, to look for the continuous symmetries associated with such a
system one has to analyse directly the equations of motion. This procedure
is more fundamental in some sense, from the point of view of symmetries,
as in certain cases many different Lagrangians may give rise to the same
equations of motion. The group analysis of the equations of motion gives
all the Lie point group symmetry generators. In the cases where a Lagrangian 
formulation is possible, a subset of these generators acting on the
Lagrangian giving zero are the Noether symmetries\cite{stephani}.
However besides these
there may be other generators obtained through group
analysis which have direct physical
significance, but not explicitly available from the consideration
of Noether symmetries alone. The reproduction
of Kepler's third law in the planetary motion problem is such an example.
This is obtained without solving the equations of motion. The extension of this
idea to the notion of Lie dynamical symmetries contains similarly a subclass
known as Cartan symmetries. The Runge-Lenz vector can be obtained from
such considerations. These symmetries are, further, related to the Lie-
B\"{a}cklund symmetries.

The application of these types of analysis to nonlocal cases have been
widely studied through B\"{a}cklund transformations and related
techniques in the context of integrable systems containing infinite
number of conservation laws\cite{ibragomov}. The Thirring model, which
corresponds to coupled partial differential equations has been analysed by
Morris \cite{morris}. The differential geometric forms developed earlier
\cite{estabrook} are used in this analysis to obtain the prolongation structure.
 Some other applications  of these ideas to important problems from 
physics is comprehensibly covered in
\cite{gaeta}.  

In this paper we restrict ourselves to finding the Lie point
symmetries of the coupled set of differential equations 
representing :

(i) a charged particle moving in three dimensions in the presence of 
a magnetic field proportional to the coordinate vector,

(ii) as in (i) and the constraint that the particle moves on the surface 
  of a sphere, the situation being equivalent to the motion of the particle
   in the presence of a magnetic monopole stationed at the centre of the
    sphere.

(iii) a particle in a particular type of velocity dependent potential.

Section II provides the outline of the method for group analysis of
the equations of motion. In section III we use this to find the generators
of the point symmetries for the above examples. Section IV is devoted to
physical interpretations and conclusions.

It should be noted that in the context of the symmetries of Wess-Zumino-Witten
 models, the ones corresponding to non Lie symmetries plays by far
the most important role and these have been fruitfully exploited  
\cite{braaten}.

\section{Group analysis}

Typically we are interested in the coupled nonlinear set
of equations representing the equations of motion of a particle in
three dimensions. These are of the form

\begin{eqnarray}
{\ddot{q^{a}}} &=& {\beta}{{\omega}^{a}} {({{q^{i}}},{\dot{q^{i}}},t )}
\end{eqnarray}
where a dot is a derivative with respect to time, $ {a,i = 1,2,}$
and  $3$,
and $ {\beta} $ is a constant involving mass,
coupling constant etc. The expressions for the function $\omega$ are given
explicitly for each example.
 
These set of equations can be analysed by means of one parameter groups 
by infinitesimal transformations. We demand the equation to be invariant under
infinitesimal changes of the explicit variable ${t}$, as well as simulteneous
infinitesimal changes of the dependent functions ${q^{a}}$ in the following
way,

\begin{eqnarray}
t \rightarrow t_1 &=&  t + \epsilon {\xi} {(t,q^1 ,q^2 ,q^3)}
 + O {({\epsilon}^2)},
\nonumber\\
{{q}^{a}} {\rightarrow} q^a_1 &=& {{q}^{a}} + {\epsilon}{{\eta}^{a} {(t,q^i)}}
 + {O {({\epsilon}^2)} } .
\end{eqnarray}

Under $t \rightarrow t_1$ and $q^a \rightarrow q^a_1$, the
equation changes to,
\begin{eqnarray}
{\ddot q}^a_1 &=& \beta \omega^a_1 (t_1, q^i_1, \dot q^i_1)
\end{eqnarray}

To illustrate the procedure consider the simple case in one space dimension.
We express the above equation in terms of ${ t} $ and ${q}$ by using the
transformation (2.2). Then the invariance condition implies that an expression
containing various partial derivatives of ${\xi}$ and ${\eta}$  is obtained
which equates to zero. For example, we get
\begin{eqnarray}
{ \frac{d{q_1}}{d{t_1}}} &=& { \frac{ dq + {\epsilon}{( {\frac{\p{\eta}}{\p{t}}} dt
+ \frac{\p{\eta}}{\p{q}}  dq )} }{ dt +
 {\epsilon}{(\frac{\p{\xi}}{\p{t}} dt
+ \frac{\p{\xi}}{\p{q}} dq )} }}   +  O ({\epsilon}^2) .
\end{eqnarray}
and now relate the left hand side with $\frac{dq}{dt} $ by using binomial
theorem for the denominator to obtain 
\begin{eqnarray}
\frac{d{q}_{1}}{d{t_1}}  &=&  \frac{dq}{dt}
 + \epsilon {( \frac{\p{\eta}}{\p{q}}
- \frac{\p{\xi}}{\p t})}{ \frac{dq}{dt}} -  {\frac{\p\xi}{\p q}}
{(\frac{\p q}{\p t})}^2  + O({\epsilon}^2)   .
\end{eqnarray}
A similar procedure is followed to express
$\frac{{d}^2 {q_1}}{{{dt}_1}^2} $
likewise. By substituting equations (2.2 ) - (2.5 ) for a given explicit expression
for $ {{\omega}_1} $ and remembering that  $ {\frac{d^2 q}{dt^2}  - \omega} $
is zero, we obtain the desired partial differential equation whose solution 
would determine $ {\xi{(t,q)}} $ and $ {\eta{(t,q)}} $. In our case, of course,
we have to find $ {\xi{(t,q^1 ,q^2 ,q^3 )}} $ and 
$ {{\eta}^a}{(t,q^1 ,q^2 ,q^3 )} $'s.
 
To relate these to the generators of the infinitesimal transformations
we write
\begin{eqnarray}
t_1 {(t, q^i ; \epsilon )} =  t +
 \epsilon {\xi (t , q^i )} +  \cdots 
= t + \epsilon{ \bf{X}} t + \cdots  \\
{{{q^a}_1} {({{t}{,}{q^i} {;}{\epsilon}}) }} = {{q^a}  +{\epsilon}
 {{\eta} {({{t}{,}{q^i}} )}} + {\cdots}}
= {{q^a} + {\epsilon} {\bf{X}}{ q^a } + {\cdots}}   
\end{eqnarray}
where the functions $\xi$ and ${\eta}^a $ are defined by
\begin{eqnarray}
{\xi (t,q^i ) } &=& {{ \frac{\p{t_1}}{\p{\epsilon}}}{{\mid}_{\epsilon = 0}}} ,\\ 
{{{\eta}^a}(t,q^i )} &=& {{\frac{\p{{q^a}_1}}{\p{\epsilon}}}{{\mid}_{\epsilon =0}}}
\end{eqnarray}
and the operator $\bf{X}$ is given by
\begin{eqnarray}
{\bf{X}} &=& {{\xi (t,q^i )} {\frac{\p}{\p{t}}}  + {{\eta}^a (t,q^i )}} {\frac{\p}
{\p{q^a}}}.
\end{eqnarray}
Following Stephani \cite{stephani}, we will consider the equation having the 
symmetry generated by $\bf{X} $ and its extension
\begin{eqnarray}
{\dot{\bf{X}}} &=& \xi {\frac{\p}{\p{ t}}} + {{\eta}^a} {\frac{\p}{{\p}{q^a}}} + 
{\dot{\eta^a}}{\frac{\p}{{\p{\dot{q^a}}}}}  
\end{eqnarray}
and the symmetry condition determines ${\dot{\eta}^a} $. The symmetry condition
is given by
\begin{eqnarray}
{\xi} {{{\omega}^a}_{,t}} + {\eta}^b {{\omega}^a}_{,b} + {( {{\eta}^b}_{,t}
+ {\dot{{q}^{c}}} {{\eta}^b}_{,c} - {\dot{q^b}} {{\xi}_{,t}}
-  {\dot{q^b}}{\dot{q^c}} {{\xi}_{,c}} )}
{ \frac{\p{{\omega}^a}}{\p{\dot{q^b}}}}\nonumber\\
+ 2 {{\omega}^a} {( {{\xi}_{,t} +  {\dot{q^b}}}{{\xi}_{,b}} )}
+ {\omega^b}{(\dot{q^a}{{\xi}_{,b}} - {{\eta}^a}_{,b}) }
+ {\dot{q^a}} {\dot{q^b}} {\dot{q^c}} {\xi}_{,bc}\nonumber\\
+ 2 {\dot{q^a}}{\dot{q^c}} {{\xi}_{,tc}}  -  {\dot{q^c}} {\dot{q^b}} {{\eta}^a}_{,bc}
+ {\dot{q^a}}{{\xi}_{,tt}} - 2 {\dot{q^b}} {{\eta}^a}_{,tb}
- {{{\eta}^a}_{,tt}}  &=& 0 
\end{eqnarray}
where $ f_{,t} = {\frac{{\p}f}{\p{t}}} $ and $ f_{,c} = {\frac{{\p}f}{\p{q^c}}} $.
By herding together coefficients of the terms with cubic, quartic, and linear 
in $ {\dot{q^a}} $ ,
and the ones  independent of $ \dot{q^a} $ separetely, and equating each of these
to zero we obtain an over determined set of partial differential equations
and solve for $ {\xi} $ and $ {{\eta}^{a}} $.

\section{The symmetry generators}

For the first example, the motion of a particle in presence
of a magnetic field, the equations of motion are 
\begin{eqnarray}
{\ddot{q^a}} &=& {\beta}{\varepsilon}^{abc}{\dot{q^b}}{q^c}
\end{eqnarray}
where the right hand side, which is our ${\omega}^a $, is the Lorentz
 force acting on a charged particle,
the magnetic field for this case being proportional to $ q^c $.

As was pointed out by Witten \cite{witten2}, one faces trouble 
in attempting to derive
these equations of motion by using the usual procedure of variation of a
Lagrangian as no obvious term can be included in the Lagrangian whose
variation would give the right hand side of equation (2.12). Hence it
would be more appropriate here to consider the group analysis of the 
equations of motion directly to obtain all the Lie point symmetries,
with the Noether symmetries being a subclass of these. However,
the present analysis cannot give any of the non-Lie symmetries.

Substituting
\begin{eqnarray}
{{\omega}^a} &=& {{{{\varepsilon}^a}_{bc}}{\dot q^b}{q^c}}
\end{eqnarray}
into equation (2.12) we obtain, in general, coupled partial differential
equations for $ \xi $ and $ \eta $ by equating to zero the terms
corresponding to various powers of $ {\dot q^l} $.

Consideration of the term with $ {\dot{q^a}}{\dot{q^b}}{\dot{q^c}} $ tells us
\begin{eqnarray}
{{\xi}_{,bc}} &=& 0   .
\end{eqnarray}
Hence we may have
\begin{eqnarray}
{\xi} &=& {{{A_l}(t)} {q^l} + B(t) + C } .
\end{eqnarray}
The terms quadratic in $ {\dot{q}} $ give
\begin{eqnarray}
{-}{\beta}{\dot{q^b}}{\dot{q^c}} {q^m}{{\xi}_{,c}}{{{\varepsilon}^{a}}_{bm}}
+ 2{\beta} {\dot{q^l}}{\dot{q^b}}{q^m}{{\xi}_{,b}}{{{\varepsilon}^{a}}_{lm}} \nonumber \\
+ {\beta} {\dot{q^r}}{\dot{q^a}}{q^s}{{\xi}_{,b}}{{{\varepsilon}^{b}}_{rs}}
+ 2 {\dot{q^a}}{\dot{q^c}}{{\xi}_{,tc}}  -  {\dot{q^c}}{\dot{q^b}}{{{\eta}^a}_{,bc}} 
&=& 0
\end{eqnarray}
This shows that $ \xi $ has to be independent of $ q^l $,
\begin{eqnarray}
{\xi} &=& B{(t)} + C  ,
\end{eqnarray}
and $ \eta $ may have the form
\begin{eqnarray}
{{\eta}^a} = { D{(t)} {q^a} + {E_l}{(t)} {{{\varepsilon}^{la}}_m} {q^m}
+  {F{(t)}} + G }  .
\end{eqnarray}
The terms linear in ${\dot q}^l$ provide
\begin{eqnarray}
  {\beta}{{\dot{q^l}}}{{{\varepsilon}^a}_{lb}} {{\eta}^{b}}
+ {\beta} {\dot{q^c}}{q^m} {{{\varepsilon}^{a}}_{bm}} {{{\eta}^b}_{,c}}
- {\beta} {\dot{q^b}}{q^m}{{{\varepsilon}^{a}}_{bm}} {{\xi}_{,t}}  \nonumber \\
+ 2{\beta}{\dot{q^l}} {q^m} {{{\varepsilon}^{a}}_{lm}}  {{\xi}_{,t}}
- {\beta} {\dot{q^r}}{q^s} {{{\varepsilon}^{b}}_{rs}} {{{\eta}^{a}}_{,b}}
- 2 {\dot{q^b}} {{{\eta}^{a}}_{,tb}}  +  {\dot{q^a}} {{\xi}_{,tt}}   
&=& 0    .
\end{eqnarray}

This demands
\begin{eqnarray}
{B{(t)}} &=& { t H  + C} ,
\end{eqnarray}
and also ${\eta}^a$ has to be independent of $t$,
\begin{eqnarray}
{{\eta}^a} &=& {-} {H} {q^a} + {E_l}{{{\varepsilon}^{la}}_{m}} {q^m}  .
\end{eqnarray}
Thus we obtain five generators
\begin{eqnarray}
{{\bf{X}}_a} &=& {{{{\varepsilon}_{a}}^{k}}_{b}} {q^b} {\frac{\p}{\p{q^k}}}\\
{{\bf{X}}_4}  &=& {\frac{\p}{\p{t}}}   \\
{{\bf{X}}_5}  &=& {t}{\frac{\p}{\p{t}}} - {q^a} {\frac{\p}{\p{q^a}}} .
\end{eqnarray}
We can find their extension from the formula
\begin{eqnarray}
{{\dot{\eta}^{a}}} = {\frac{d{{\eta}^a}}{dt}} - {\dot{q^a}}{\frac{d{\xi}}{dt}}
\end{eqnarray}
and obtain,with the extensions
\begin{eqnarray}
{\dot{{\bf{X}}_a}}  &=&  {{{{\varepsilon}_{a}}^{k}}_{b}} {( {q^b}{\frac{\p}{\p{q^k}}}
+  {\dot{q^b}}{\frac{\p}{\p{\dot{q^k}}}} )}  \\
{\dot{{\bf{X}}_4}} &=&  {\frac{\p}{\p{t}}}  \\
{\dot{{\bf{X}}_5}}  &=&  { t} {\frac{\p}{\p{t}}} - {q^a}{\frac{\p}{\p{q^a}}}
- 2 {\dot{q^a}} {\frac{\p}{\p{\dot{q^a}}}} .
\end{eqnarray}

A comparison with the results of similar analysis for the Kepler problem
\cite{stephani} shows that the first four generators are identical, the first
three corresponding to the generators of the three dimensional rotation group
and $ {{\bf{X}}_4} $ is the generator for time translation. However in this case
the law corresponding to Kepler's third law goes instead like
\begin{eqnarray}
{{t}_1} {{r}_1} &=& tr
\end{eqnarray}

If the particle is further constrained to move on the surface of a sphere of
unit radius, the equation of motion becomes
\begin{eqnarray}
{\ddot{q^a}} &=& {\beta}{{{\varepsilon}^{a}}_{bc}} {\dot{q^b}}{\dot{q^c}}
- {q^a}{\dot{q^k}}{\dot{q_k}}  .
\end{eqnarray}
This is equivalent to the case of a particle moving in the presence of
a magnetic monopole centered at the origin of the sphere. Witten has 
generalised this idea to arbitrary dimensions for field theoretic 
considerations. 

With $ {\omega}^a $ being equal to the right hand side of 
equation (3.31)
the group analysis shows that there is only one trivial 
time translation
generator for this problem,
\begin{eqnarray}
{\bf{X}} &=& {\frac{\p}{\p{t}}}  .
\end{eqnarray}
Same is the case if we ignore the term
containing ${\varepsilon}^{abc}$in equation (3.31).

But for equations of motions of the form
\begin{eqnarray}
{\ddot{q^a}} &=& {\dot{q^a}}{q^k}{q_k}  
\end{eqnarray}
we again find five symmetry generators, the first four being the same
as ${\dot{\bf{X}}}_a $ and $ {\dot{\bf{X}}}_4 $ while
\begin{eqnarray}
{\bf X}'_5 &=& 2 t {\frac{\p}{\p{t}}}
 - {q^a}{\frac{\p}{\p{q^a}}}  ,
\end{eqnarray}
and with its extension,
\begin{eqnarray}
{\dot{\bf X}}'_5 &=& 2 t {\frac{\p}{\p{t}}} - {q^a}{\frac{\p}{\p{q^a}}}
- 3 {\dot{q^a}}{\frac{\p}{\p{\dot{q^a}}}} .
\end{eqnarray}
The length and time scale in this case as
\begin{eqnarray}
{t_1}{{r_1}^2} &=& t{r^2} .
\end{eqnarray}

\section{Conclusion}

It is well known that if the right hand side of equation (3.13) is zero,
it would admit eight symmetries, which is the maximum number for an
ordinary second order differential equation. By including different
$ {q^a} $, $ {\dot{q^a}} $ dependent terms in the equations we
do explicitly see which generators survive and also can 
find the complete
Lie algebra. We have chosen three cases motivated by problems from
physics. The original motivation of including Wess-Zumino terms
in the Lagrangian has been to reduce some of its symmetries\cite{witten2},
but as we see in three dimensions the effect is rather drastic
for the point symmetries. Only the trivial time translation generator
survives. However, for the other examples considered we get some interesting
result in the form of Kepler's scaling law. This we get without solving
the equations of motion. Besides this, if a Langrangian could be set up,
those generators operating on the Lagrangian to produce zero include
all the Noether symmetries\cite{stephani}.

There has been much current interest in the gauge theories on noncommutative
spaces in connection with the quantization of D-branes\cite{bigatti}.
 There the models
considered are usually in two dimensional space and relates to the phenomena
of quantum Hall effect as the vector potential taken is proportional
to the cordinates. This gives rise to the necessary 
constant (strong) magnetic field.
However, in our case the setting is in three space dimensions and
it is the magnetic field which is proportional to the coordinate
vectors. 

It is also expected that related analysis may provide useful information
when terms are modified in the Lagrangian, due to quantum corrections
for example. By explicitly showing how many and which generators remain
as symmetries we will have a better understanding of the breaking
of continuous symmetries.

\noindent{\bf Acknowledgement}
 
I would like to thank the Abdus Salam International Centre for 
Theoretical Physics, Trieste for the kind hospitality where 
this work was done.



\begin{thebibliography}{99}

\bibitem{wess} J. Wess and B. Zumino, Phys. Lett. {\bf B37}, 95 (1971).

\bibitem{witten} E. Witten, Nucl. Phys. {\bf {B 223}}, 422 (1983).

\bibitem{stephani} H. Stephani, {\it Differential equations,
their solution
using symmetries}, Cambridge University Press, Cambridge,(1989); \\
P. J. Olver, {\it Application of groups to differential equations}, 
Springer, Berlin (1986); \\
L. V. Ovsiannikov, {\it Group analysis of differential equations}, 
Academic Press, New York (1982).

\bibitem{ibragomov} N. H. Ibragimov,(ed.) {\it CRC handbook
of Lie group analysis of differential equations},
CRC Press, Boca Raton (1994).

\bibitem{morris} H. C. Morris, J. Phys. A {\bf{12}}, 131 (1979); \\
D. David, J. Math. Phys. {\bf {25}}, 3424(1984).

\bibitem{estabrook} H. D. Wahlquist and F. B. Estabrook,
J. Math. Phys. {\bf{16}}, 1 (1975).

\bibitem{gaeta} G. Gaeta, {\it Nonlinear symmetries and
nonlinear equations},
Kluwer Academic Publishers, Dodrecht, (1994).

\bibitem{braaten} E. Braaten, T. L. Curtright and C. K. Zachos, 
Nucl. Phys {\bf B260}, 630 (1985); \\
M. Forger and P. Zizzi, Nucl. Phys {\bf B287}, 131 (1987).

\bibitem{witten2} E. Witten, Commun. Math. Phys {\bf{92}}, 455 (1984);\\
W. J. Zakrzeweski, {\it Low dimensional sigma models},
Adam Hilger,Bristol,(1989).


\bibitem{bigatti} D. Bigatti and L. Susskind, Phys. Rev.
{\bf{D62}}, 066004 (2000) [hep-th/9908056].

\end{thebibliography}
\end{document}